\documentclass{article}
\begin{document}

\begin{center}
{\Large A Simple Path Integration For the Time Dependent Oscillator}
\end{center}
\vspace{5mm}
\begin{center}
H. Ahmedov\footnote{Feza G\"ursey Institute, P.O. Box 6, 81220,
\c{C}engelk\"{o}y, Istanbul, Turkey, E-mail: hagi@gursey.gov.tr}, I.
H. Duru\footnote{Turkish Academy of Sciences (TUBA), E-mail:
duru@gursey.gov}, and A.E. Gumrukcuoglu \footnote{Feza G\"ursey
Institute, P.O. Box 6, 81220,  \c{C}engelk\"{o}y, Istanbul, Turkey }
\end{center}

\begin{center}
{\bf Abstract}
\end{center}
Feynman propagator is calculated for the time dependent harmonic
oscillator by converting the problem into a free particle motion

\vspace{1cm} {\bf I. Introduction}

One way of treating the time dependent oscillator described by the
Hamiltonian
\begin{equation}\label{1}
   H=\frac{p^2}{2\mu}+\frac{\mu}{2}\omega^2(t)q^2
\end{equation}
is employment of the time-dependent point canonical transformations
of type \cite{1}
\begin{eqnarray}\label{2}
  x &=& f(t) Q \nonumber \\
  p &=& \frac{1}{f(t})P
\end{eqnarray}
generated by $$ F_2(q,P,t) =\frac{qP}{f(t)}.$$

The above transformation has been used to obtain the Green function
by choosing $f(t)$ to map the problem into the usual oscillator with
constant frequency $\omega^2_0$.

Another method of approaching the problem is the making use of the
invariants. The Hamiltonian (\ref{1}) is known to possess a
time-dependent invariant given by \cite{2}
\begin{equation}\label{4}
  I=\frac{1}{2}(\dot{q}p-q\dot{p})^2+\frac{\omega^2_0}{2}(\frac{q}{p})^2,
  \ \  \omega^2_0= constant
\end{equation}
The above constant $\omega^2_0$ and the constant frequency of the
usual oscillator into which the original problem mapped by (\ref{2})
are in fact the same. The treatment which uses the invariant
approach usually find the wave functions as well as Feynman
propagators again by mapping the problem into the constant frequency
ones \cite{3}.

In this note we will show that a more simpler choice for the
function $f(t)$ of (\ref{2}), maps the path integration for the
time-dependent original problem into the free motion path integral
in $Q, P$ - phase space. In other words the choice we employed for
$f(t)$, corresponds to the employment of time-dependent invariant
(\ref{4}) with  $\omega_0=0$. In fact such a choice for $\omega_0$
has been introduced for writing a general expression for  the wave
function corresponding to (\ref{1}) which is of the free wave
function form \cite{4}. However it has, to our knowledge not been
adopted in path integrations.

In the next section we briefly summarize the employment of the
transformations (\ref{2}) in the path integration for the problem,
which is essentially same as the one studied in \cite{1}. We then
make our choice for $f(t)$, and write the general expression for the
Kernel.

In the last section we simply present the results for some
interesting forms of the frequency $\omega (t)$.

\vspace{1cm} {\bf II. Path Integration with Time-Dependent Point
Canonical Transformations}

The probability amplitude for the particle to move from the
space-time point $q_a, t_a$ to $q_b, t_b$ under the influence of the
time dependent oscillator potential is
\begin{equation}
K(q_a,t_a; q_b, t_b)=\int DqDp e^{i\int_{t_a}^{t_b} dt
(p\dot{q}-\frac{p^2}{2\mu}-\frac{\mu}{2}\omega^2(t)q^2)}
\end{equation}
which in its explicit time-sliced form expressed as
\begin{equation}\label{6}
K(q_a,t_a; q_b,
t_b)=\lim_{n\rightarrow\infty}(\prod_{j=1}^{n}\int_{-\infty}^{\infty}dq_j)
(\prod_{j=1}^{n+1}\int_{-\infty}^{\infty}\frac{dp_j}{2\pi})
\prod_{j=1}^{n+1} e^{i[ p_j(q_{j}-q_{j-1})-\frac{\varepsilon
p_j^2}{2\mu}-\frac{\varepsilon\mu}{2}\omega^2(t_j)q_j^2)] }
\end{equation}
with
\begin{equation}
    t_j=j\varepsilon, \ \ \  (t_b-t_a) = (n+1) \epsilon \nonumber \\
\end{equation}
and
\begin{equation}
    q_a, t_a = q_0, t_0; \ \ \ q_b, t_b = q_{n+1}, t_{n+1}. \nonumber
    \\
\end{equation}
Point canonical transformations (\ref{2}) transforms Hamiltonian and
the Action as
\begin{equation}
 H_Q= H+ \frac{\partial F_2}{\partial F_2}=
 \frac{P^2}{2\mu
 f^2(t)}+\frac{\mu}{2}\omega^2(t)f^2(t)Q^2-\frac{\dot{f}(t)}{f(t)}QP
\end{equation}
and
\begin{equation}
\int_{t_a}^{t_b} dt (p\dot{q}-H)=\int_{t_a}^{t_b} dt (P\dot{Q}-H_Q).
\end{equation}
The path integral measure of (\ref{6}) transforms as
\begin{equation}
   \frac{dp_{n+1}}{2\pi}
   \prod_{j=1}^{n}\int_{-\infty}^{\infty}\frac{dq_jdp_j}{2\pi}=
   \frac{dP_{n+1}}{2\pi f(t_{n+1})}
   \prod_{j=1}^{n}\int_{-\infty}^{\infty}\frac{dQ_jdP_j}{2\pi}.
\end{equation}
The factor $f^{-1}(t_{n+1})=f^{-1}(t_b)$ can be symmetrized as
\begin{equation}
\frac{1}{f(t_b)}=\frac{1}{\sqrt{f(t_a)f(t_b)}}e^{-i\frac{1}{2}\ln
\frac{f(t_b)}{f(t_a)}}.
\end{equation}
Note that we could have arrange the time-slicing from $j=0$ to
$j=n$; in which case we would have an extra momentum integration
over $dp_0$ instead of $dp_{n+1}$ in (\ref{6}). The symmetrization
procedure would lead the same expression as (10)  with the sign in
the exponent is reversed. Taking the average of these two
time-slicing recipes removes the exponential.

The path integration (5) is, after  the employments of
transformations given by (7) and (8) and the symmetrization (10)
becomes ( after making a translation $P\rightarrow P+
\dot{f}(t)f(t)\mu $ )
\begin{equation}
K(q_a,t_a; q_b, t_b)=\frac{1}{\sqrt{f(t_a)f(t_b)}}
e^{\frac{i\mu}{2}(\dot{f}(t_b)f(t_b)Q_b^2-\dot{f}(t_a)f(t_a)Q_a^2)}
\tilde{K}(Q_a,t_a;Q_b, t_b)
\end{equation}
where
\begin{equation}
\tilde{K}(Q_a,t_a;Q_b, t_b)=\int DQDP e^{i\int_{t_a}^{t_b} dt
(P\dot{Q}-\frac{P^2}{2\mu
f^2(t)}-\frac{\mu}{2}(\omega^2(t)f^2(t)+\dot{f}(t)f(t) )Q^2)}
\end{equation}
Up to this point we have summarized the known procedure \cite{1}.
Now we make the choice for $f(t)$ to satisfy simple equation
\begin{equation}
  \ddot{f}+\omega^2(t)f=0
\end{equation}
which is same as the one previously employed in studying the
invariant (3)  to write a general form for the wave function
\cite{4}. With (14) the kernel (13) takes the free propagator form
and this admits the simple solution:
\begin{equation}
\tilde{K}(Q_a,t_a;Q_b, t_b)=\sqrt{\frac{\mu}{2i\pi W}}
e^{i\frac{\mu}{2W}(Q_b- Q_a)^2}
\end{equation}
where
\begin{equation}
W=\int_{t_a}^{t_b} \frac{dt}{f^2(t)}.
\end{equation}
Inserting (14) into (11) we arrive at the final form for the
original Kernel:
\begin{equation}
K(q_a,t_a; q_b, t_b)=\sqrt{\frac{\mu}{2i\pi f(t_a)f(t_b)W}}
e^{i\frac{\mu}{2}(
\frac{\dot{f}(t_b)}{f(t_b)}q_b^2-\frac{\dot{f}(t_a)}{f(t_a)}q_a^2)}
e^{i\frac{\mu}{2W}(\frac{q_b}{f(t_b)}- \frac{q_a}{f(t_a)})^2}
\end{equation}
It is strait forward to verify that the above expression is indeed
the correct Green function for the Hamiltonian (1).

\vspace{1cm} {\bf III. Examples and Conclusion }

For a given frequency all one has to do is to solve the Riccati
equation (14) and then insert the solution into the "free" Kernel in
(17)
\\
Some examples are the following
\\
\\
(i) $\omega(t) = \omega_0 = constant$, $$f(t)=\cos\omega_0t$$
\\
\\
(ii) An exponentially decaying frequency: $\omega^2(t) = \omega^2_0
e^{-\alpha t}$; $\omega_0, \alpha = constants$. $$f(t) =
J_0(\frac{2\omega_0}{\alpha}e^{-\alpha t/2})$$.
\\
\\
(iii) $\omega^2(t) = (\omega_0\alpha^\beta)^2 t^\beta$; $\omega_0,
\alpha, \beta$ are arbitrary constants ( i.e., not necessarily
integer power ).
$$f(t) = \sqrt{t/\omega_0\alpha^\beta} J_{\frac{1}{\beta+2}}
(\frac{2\omega_0\alpha^\beta}{\beta + 2} t^{\frac{\beta+2}{2}})$$.
\\
\\
(iv) A $\delta-$ function pulse frequency given by
$$ \omega^2(t)=\omega^2_0 \delta (t-t_0) + \omega^4_0\theta
(t-t_0);  \ \omega_0, t_0 = constants$$ the function $f(t)$ is
$$f(t)= e^{\omega_0|t-t_0|}.$$
\\
\\
(v) A broader time dependence around $t_o$ may be given by
$$\omega^2(t)=\frac{\alpha^2}{\cosh^2\beta (t-t_0)}; \ \alpha, \beta = constants$$
is solved for
$$f(t) = P_{i\sqrt{\alpha^2-1/4}-1/2} (\tanh \beta (t-t_0))$$
where $P$ is a Legendre function.

\vspace{1cm} {\bf Acknowledgments} The authors  H. Ahmedov and I. H.
Duru thanks the Turkish Academy of Sciences (TUBA) for its support.

\end{document}